\documentclass[9pt,twocolumn,twoside]{osajnl}
\usepackage{graphicx}
\usepackage{subfigure}
\usepackage{dcolumn}

\journal{optica} 

\setboolean{shortarticle}{false}

\title{Breaking Rayleigh's Criterion via Discernibility in High-Dimensional Light-Field Space with Snapshot Ghost Imaging}

\author[1,2, \dag]{Zhishen Tong}
\author[1,\dag]{Zhentao Liu}
\author[3,4]{Jian Wang}
\author[1]{Xia Shen}
\author[1,5,*]{Shensheng Han}

\affil[1]{Key Laboratory for Quantum Optics and Center for Cold Atom Physics of CAS, Shanghai Institute of Optics and Fine Mechanics, Chinese Academy of Sciences, No. 390 Qinghe Road, Jiading District, Shanghai, Shanghai, 201800, China}
\affil[2]{Center of Materials Science and Optoelectronics Engineering, University of Chinese Academy of Sciences, Beijing, 100049, China}
\affil[3]{School of Data Science, Fudan University, Shanghai, 200433, China}
\affil[4]{ZJLab, Fudan-Xinzailing Joint Research Centre for Big Data, Shanghai Key Lab of Intelligent Information Processing, Shanghai, 200433, China}
\affil[5]{Hangzhou Institute for Advanced Study, University of Chinese Academy of Sciences, Hangzhou, 310024, China}
\affil[$\dag$] {These authors equally contributed to this work.}
\affil[*]{Corresponding author: sshan@mail.shcnc.ac.cn}

\begin{abstract}

By encoding the high-dimensional light-field imaging information into a detectable two-dimensional speckle plane, ghost imaging camera via sparsity constraints (GISC camera) can directly catch the high-dimensional light-field imaging information with only one snapshot. This makes it worth to revisit the spatial resolution limit of this optical imaging system. In this paper we show both theoretically and experimentally that, while the resolution in high-dimensional light-field space is still limited by diffraction, the statistical spatial resolution of GISC camera can be greatly improved comparing to classical Rayleigh's criterion by utilizing the discernibility in high-dimensional light-field space.
The interaction between imaging resolution, degrees of freedom of light field, and degrees of freedom of objects in high-dimensional light-field space is also demonstrated.
\end{abstract}

\begin{document}

\maketitle

\section{Introduction}
Over a century, Rayleigh's criterion has been the most influential resolution limit for incoherent optical imaging systems~\cite{born2013principles}. In recent decades, various optical techniques breaking Rayleigh's criterion have emerged. By utilizing structured light, one can break the diffraction limit of the corresponding imaging system, represented by the structured illumination microscopy (SIM)~\cite{heintzmann2009subdiffraction} and Fourier ptychographic microscopy~\cite{zheng2013wide}, despite the fact that they are still limited by diffraction of whole imaging equipment including illumination system. While for scanning imaging systems, reducing the size of scanning light spot based on near-field or stimulated emission depletion effect, such as the near-field scanning optical microscopy (NSOM)~\cite{durig1986near} and stimulated emission depletion microscopy (STED)~\cite{hell1994breaking}, also render Rayleigh's criterion irrelevant to the imaging resolution. Another category for the far-field super-resolution technique is mainly based on statistical methods, including the stochastic optical reconstruction microscopy (STORM)~\cite{rust2006sub} and photoactivated localization microscopy (PALM)~\cite{betzig2006imaging}, where the positions of fluorescent  molecules are estimated and then a high-resolution image as a whole is formed. More recently, fundamental resolution limit based on quantum properties of light field for classical and quantum light sources have been also investigated to overcome ``Rayleigh's curse''~\cite{tsang2016quantum,tsang2018conservative}. 
In parallel to above physical super-resolution techniques, plenty of super-resolution algorithms~\cite{chaudhuri2001super} are developed to create super-resolution image from low-resolution images, such as the spectrum extrapolation~\cite{papoulis1975new}, multi-frame super-resolution~\cite{park2003super}, and super-resolution image via deep learning~\cite{dong2015image}. 

\begin{figure*}[t] 
\centering   
\subfigure[~~Schematic of GISC camera]  
{\includegraphics[width=\columnwidth]{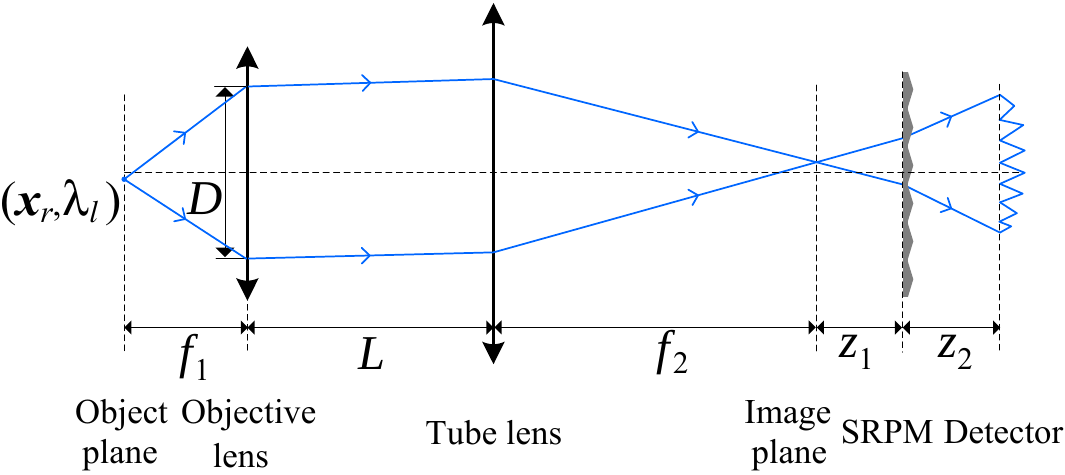}  
 \label{fig:a}   
}   
\subfigure[~~Statistical resolution limit of GISC camera~($K = 2, 3, 5, 7$). ]{ 
\label{fig:b}     
\includegraphics[width=\columnwidth, height=42mm]{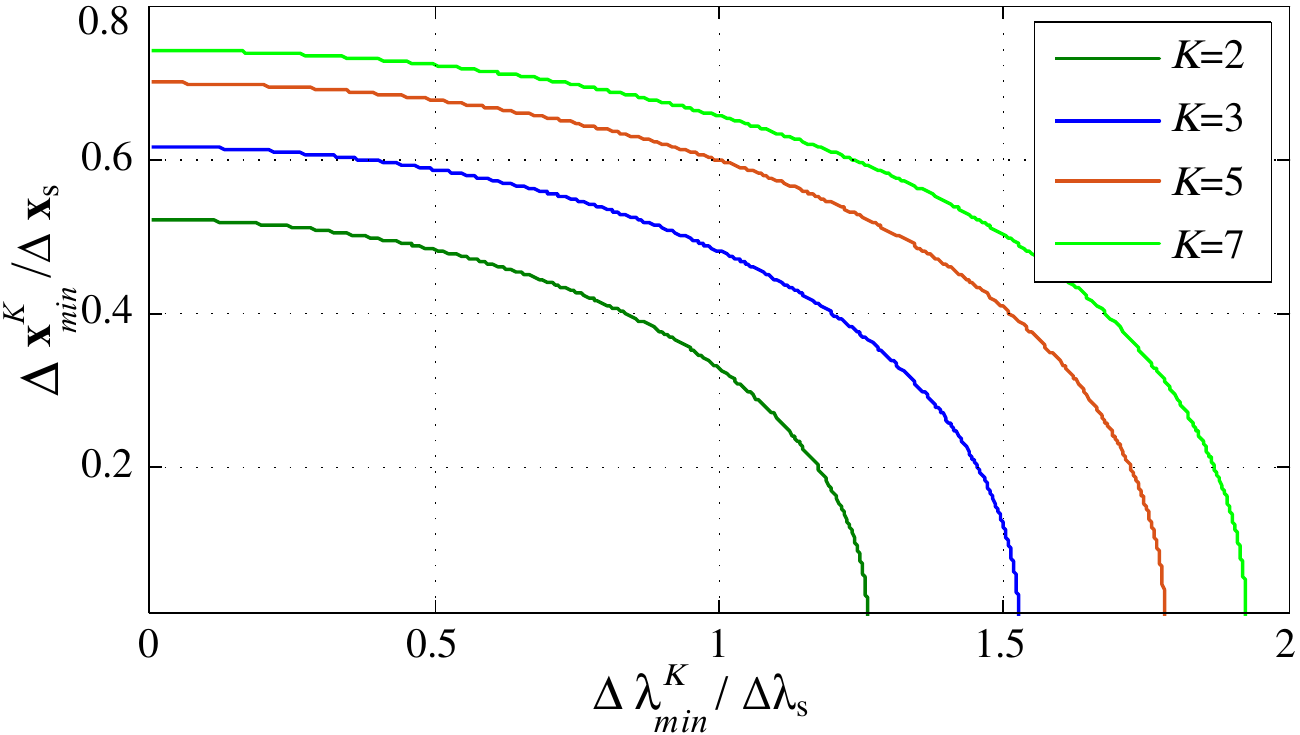}     
}   
 \vspace{-3mm}
\caption{(a)~GISC camera's schematic. SRPM, spatial random phase modulator; (b) Statistical resolution limit of GISC camera in ($\mathbf{x}, \lambda$) light-field space, $\Delta \mathbf{x}^{K}_{min}$ and $\Delta \lambda^{K}_{min}$ are the nearest resolved spatial distance and spectral distance among $K$ points respectively, $\Delta \mathbf{x}_s$ is Rayleigh's  spatial resolution criterion, and $\Delta \lambda_s$ is the spectral resolution of GISC camera.} 
\vspace{-3mm}    
\end{figure*}

Light field can be fully characterized by a $7$-dimensional plenoptic function, including the spatial, spectral, temporal and angular dimensions~\cite{adelson1991plenoptic}. Shannon's channel capacity of an imaging system is $\mathcal{H} = N_{DoF} log(1+SNR)$, which is determined by the detection signal-to-noise ratio ($SNR$) and all degrees of freedom of light field $N_{DoF}=N_t \times N_s \times N_c \times N_{\phi}$, where $N_t, N_s, N_c, N_{\phi}$ are the temporal, spatial, spectral and polarization degrees of freedom, respectively~\cite{Toraldo1955resolving,Toraldo1969degreesofimage,liu2016spectral}.
So far, almost all imaging systems have been designed based on the ``point-to-point" imaging mode, where the emission light from each point in real space (or spacial Fourier space) on the object plane is directly recorded by the corresponding pixel of a detector on the image plane. Thus, the super-resolution ability of this imaging mode can be understood as increasing the precision of separating two adjacent points in real space. This type of imaging mode, however, is difficult to directly imaging in high-dimensional light-field space with a two-dimensional (2D) photon detector, and thus ``degenerates'' degrees of freedom of light field other than spatial degrees of freedom $N_s$, making it hard to fully utilize the imaging channel capacity. Different from the traditional imaging mode, ghost imaging (GI) extracts the object's information from the mutual correlation of the light-field intensity fluctuation~\cite{erkmen2010ghost}. By encoding the high-dimensional information of light field irradiating from object into speckle patterns on a detectable 2D plane, ghost imaging camera via sparsity constraints (GISC camera)~\cite{liu2016spectral} can simultaneously obtain object's high-dimensional information with one snapshot. Degrees of freedom of light field in GISC camera are no longer ``degenerate''.
Thus, besides directly enhancing the accuracy of determining positions of two point sources in real space, GISC camera provides new possibility to break Rayleigh's spatial resolution criterion via the discernibility in high-dimensional light-field space.


In this paper,  we demonstrate that while the resolution of GISC camera in high-dimensional light-field space is still restricted by diffraction, the spatial resolution can be greatly improved by exploiting the discernibility in other dimensions of the light-field space, such as the spectrum and polarization. Specifically, we analytically quantify the interaction between imaging resolution, degrees of freedom of light field, and degrees of freedom of object. The relationship between the spatial super-resolution ability of GISC camera and discernibility in high-dimensional light-field space is also investigated experimentally. 
\section{Theory}

GISC camera utilizes a spatial random phase modulator (SRPM) after the image plane of traditional imaging system, such as the microscope and telescope, to modulate the light field into speckle field. As shown in Figure~\ref{fig:a}, a typical traditional imaging system consists of the objective lens with focus length $f_1$ and aperture $D$, and tube lens with focus length $f_2$. The distance between two lens is $L$. An SRPM is placed after the image plane with distance $z_1$, and a detector is placed behind SRPM with distance $z_2$.

\begin{figure*}[t]
\centering
\includegraphics[width=0.83\linewidth]{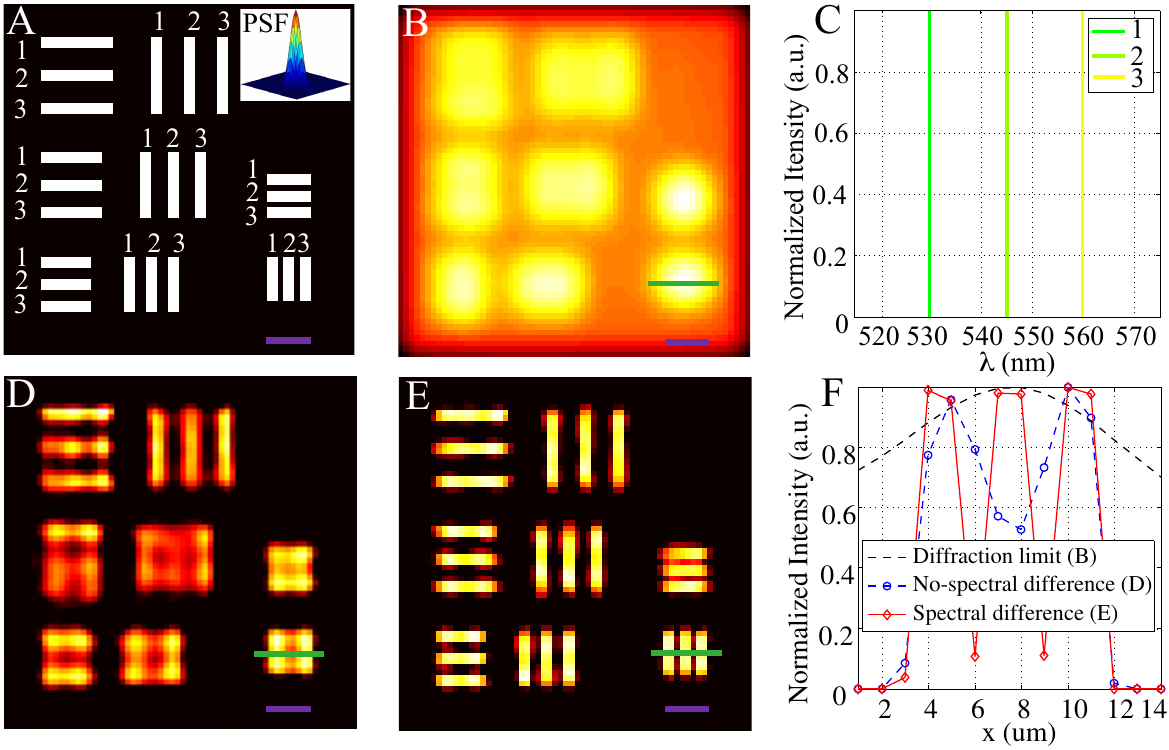}
\vspace{-3mm}
\caption{Simulation results. (A) Resolution test target, the width of each slit is $2$ um, and the spatial distances between each slit are respectively $6$ um, $5$ um, $4$ um and $3$ um; the purple scale bar is $8$ um, corresponding to the FWHM of PSF of the objective lens in GISC camera. (B) Diffraction limited image. (C) Wavelength of each slit labelled $1, 2, 3$ on resolution test target (A), when the resolution test target has spectral difference. (D) Reconstructed result of resolution test target without spectral difference. (E) Reconstructed result of resolution test target with spectral difference. (F) Comparison of resolution enhancement. Intensity profiles extract from the cross-section green lines in (B), (D) and (E).}
\label{fig:SimulationResults}
\vspace{-3mm}
\end{figure*}

Considering an incoherent illumination, the speckle intensity on the detection plane of GISC camera is~\cite{liu2016spectral}
\begin{equation}
I_t(\mathbf{x}_t) = \int \hspace{-2mm} \int T( \mathbf{x}_r, \lambda_l) I_r ( \mathbf{x}_t; \mathbf{x}_r, \lambda_l ) d \mathbf{x}_r d \lambda_l, \label{eq:It}
\end{equation}
where $T(\mathbf{x}_r, \lambda_l)$ denotes the image of object in the spatial and spectral dimensions ($\mathbf{x}_r, \lambda_l$), $I_r(\mathbf{x}_t; \mathbf{x}_r, \lambda_l)$ is the system response of GISC camera corresponding to a monochrome point light source with wavelength $\lambda_l$ at the position $\mathbf{x}_r$, which is detected in the pre-determined process, and $ \mathbf{x}_r $ and $ \mathbf{x}_t$ are 2D vectors on the object plane and detection plane of GISC camera, respectively.

From the viewpoint of GI~\cite{cheng2004incoherent}, each pixel of the detector in GISC camera acts as a bucket detector at the test arm, and $I_r(\mathbf{x}_t; \mathbf{x}_r, \lambda_l)$  at the fixed $\mathbf{x}_t$ for different positions $\mathbf{x}_r$ and wavelength $\lambda_l$ serves as a pixelated signal at the reference arm. The image of object in the spatial and spectral dimensions is recovered by calculating the second-order correlation function between the intensity fluctuations at the pre-determined reference arm and the test arm,
\begin{eqnarray}
\Delta G^{( 2,  2  )} (  \mathbf{x}_r,  \lambda_l ) &\hspace{-2mm}=& \hspace{-2mm}   \langle  \Delta I_t( \mathbf{x}_t)    \Delta I_r(  \mathbf{x}_t;  \mathbf{x}_r, \lambda_l ) \rangle_{\mathbf{x}_t}  \nonumber\\
& \hspace{-28mm} \propto & \hspace{-15mm} \int  \hspace{-2mm} \int T  (  \mathbf{x}'_r,  \lambda'_l  )g^{(2)}(\mathbf{x}_r,\mathbf{x}'_r; \lambda_l, \lambda'_l) d \mathbf{x}'_r  d\hspace{-.05mm}  \lambda'_l  \nonumber\\
& \hspace{-28mm} = & \hspace{-15mm} T  (  \mathbf{x}_r,  \lambda_l  ) \hspace{-.25mm}\otimes \hspace{-.25mm} \exp\hspace{-1mm}\left [ \hspace{-.5mm} { \hspace{-.25mm} - \hspace{-0.25mm}  \left( \hspace{-.25mm}  \frac{ 2 \pi ( n \hspace{-0.25mm} -\hspace{-0.25mm}  1 \hspace{-0.25mm} )  \omega \lambda_l  }{ {\overline{ \lambda }} ^2} \hspace{-.5mm} \hspace{-.5mm} \right)^2 } \hspace{-.25mm} \right ] \hspace{-1mm} \hspace{-.5mm}\left( \hspace{-.5mm} \frac{ 2 \hspace{-0.25mm}J_1 \hspace{-0.25mm}( \hspace{-0.25mm}  \frac{ \pi D \mathbf{x}_r}{ \overline{ \lambda } f_1 }  \hspace{-0.25mm}) } { \frac{ \pi D \mathbf{x}_r}{  \overline{ \lambda } f_1 }}\hspace{-.5mm} \right)^2,~~~~~\label{eq:deltag2}
\end{eqnarray}
where 
\begin{small}
\begin{eqnarray} 
g^{(2)}( \hspace{-.25mm} \mathbf{x}_r, \hspace{-.25mm} \mathbf{x}'_r; \hspace{-.25mm} \lambda_l, \hspace{-.25mm} \lambda'_l \hspace{-.25mm} )  &\hspace{-2.5mm} = &\hspace{-2.5mm} \exp  \hspace{-1mm} \left [ \hspace{-.5mm} { -  \hspace{-1mm} \left( \hspace{-.25mm}  \frac{ 2 \pi ( n \hspace{-0.25mm} -\hspace{-0.25mm}  1 \hspace{-0.25mm} )  \omega |\lambda_l \hspace{-0.25mm} - \hspace{-0.25mm} \lambda'_l | }{ {\overline{ \lambda }} ^2} \hspace{-.5mm} \hspace{-.25mm} \right)^2 } \hspace{-.25mm} \right ] \hspace{-1.5mm} \left( \hspace{-.5mm} \frac{ 2 \hspace{-0.25mm}J_1 \hspace{-0.25mm}( \hspace{-0.25mm}  \frac{ \pi D |\mathbf{x}_r \hspace{-0.25mm} - \hspace{-0.25mm} \mathbf{x}'_r|}{ \overline{ \lambda } f_1 }  \hspace{-0.25mm}) } { \frac{ \pi D |\mathbf{x}_r \hspace{-0.25mm} - \hspace{-0.25mm} \mathbf{x}'_r |}{  \overline{ \lambda } f_1 }}\hspace{-.5mm} \right)^2, \nonumber~~~~
\end{eqnarray}
\end{small}and $\langle \cdot \rangle_{\mathbf{x}_t} $ means the spatial average over the coordinate $\mathbf{x}_t$,  $\otimes$ denotes the convolution operator, $\overline{\lambda}$ is the center wavelength, $J_1(\cdot)$ is the 1st-order Bessel function, $n$ is the refractive index of SRPM, and $\omega$ is the standard deviation of the height of SRPM. \eqref{eq:deltag2} holds under the condition
$
\frac{ \zeta | z_1 + z_2 |}{ \omega | z_2 | } \ll 2 \pi ( n - 1 )\frac{f_1}{D}
$;
 see Supplementary material for more detailed derivation. The normalized second-order correlation function $ g^{(2)}(\mathbf{x}_r,\mathbf{x}'_r; \lambda_l, \lambda'_l) $ in~\eqref{eq:deltag2} represents the correlation of system's responses of two points ($\mathbf{x}_r, \lambda_l$) and ($\mathbf{x}'_r, \lambda'_l$), and generally characterizes the conventional resolution of GI~\cite{cheng2004incoherent}.

~\eqref{eq:It} can be also rewritten in the matrix form as~\cite{liu2016spectral}
\begin{equation}\label{sample}
\mathbf{Y} = \mathbf{\Phi} \mathbf{X},
\end{equation}
where $\mathbf{Y} \in \mathbb{R}^m$ are the sampling signals composed of the speckle intensity $I_t(\mathbf{x}_t)$, the sampling matrix $ \mathbf{\Phi} \in \mathbb{R}^{ m \times n} $ whose columns consist of the speckle patterns $I_r ( \mathbf{x}_t; \mathbf{x}_r, \lambda_l )$, $m$ is the number of samples, and $\mathbf{X} \in \mathbb{R}^{n}$ denotes the image of object in the spatial and spectral dimensions. Inspired by the prevalent compressive sensing (CS) theory~\cite{donoho2006compressed, candes2005decoding}, which exploits the sparsity prior, $\mathbf{X}$ can be retrieved by solving the problem
\begin{equation}
\mathop {\min }\limits_\mathbf{X} {\left\| \mathbf{X} \right\|_0}~~ \rm{subject~ to}~~~~ \mathbf{Y} - \bar{ \mathbf{Y} }  = ( \mathbf{\Phi} - \bar { \mathbf{\Phi} } ) \mathbf{X},
\label{equation: measurement}
\end{equation}
where for matrix $\mathbf{A}\in \mathbb{R}^{ m \times n} $, $ \bar{ \mathbf{A} }_{ i j} =\frac{1}{m} \sum_{k= 1}^{m} \mathbf{A}_{ k j}, \forall j \in \{1,\cdots, n \}$. For notational simplicity, denote $\mathbf{\Psi} = \mathbf{\Phi} - \bar{ \mathbf{\Phi} }$.


In CS, the mutual coherence of matrix $\mathbf{\Psi}$ has been commonly used to analyze the performance of algorithms~\cite{donoho2001uncertainty}. It is defined as
\begin{equation}
\mu (\mathbf{\Psi}) := \mathop {\max}\limits_{1 \leq i < j \leq n} \frac{  |  \psi^{T}_i \psi_j | } { \| \psi_i \|_2  \| \psi_j \|_2 },
\label{mutual coherence}
\end{equation}
where $\|\cdot\|_2$ is the $\ell_2$-norm, and $\psi_i$ denotes the $i$-th column of $\bf{\Psi}$.  
The relationship between the normalized second-order correlation function $g^{(2)}(\mathbf{x}_r,\mathbf{x}'_r; \lambda_l, \lambda'_l)$ and the mutual coherence $\mu(\mathbf{\Psi})$ can be given as
\begin{eqnarray}
\hspace{-2.5mm} \mu (\mathbf{\Psi})  & \hspace{-2mm} =  & \hspace{-2mm}  \mathop{ \max } \limits_{ 1 \leq i < j \leq n } \frac{ \left| ( \phi_i - \bar{ \phi_ i } )^ T ( \phi_j - \bar{ \phi_j }) \right| } { \| \phi_i - \bar{ \phi_ i } \|_2  \| \phi_j - \bar{ \phi_j } \|_2 } \nonumber \\
& \hspace{-1mm} \overset{(a)} {\approx} & \hspace{-1mm}  \mathop{ \max } \limits_{1 \leq i < j \leq n} \frac{   \left| \phi^T_i \phi_j  - m \langle \phi_ i \rangle  \langle \phi_ j \rangle  \right| } { \sqrt{ m C( \phi_i) } \langle \phi_ i \rangle \sqrt{ m C( \phi_j) } \langle \phi_ j \rangle }  \nonumber \\
&\hspace{-2mm} = &\hspace{-2mm}  \mathop{ \max } \limits_{1 \leq i < j \leq n } \left| \frac{ \langle \phi_i \phi_j \rangle } { \langle \phi_i \rangle \langle \phi_j \rangle  } - 1 \right | \hspace{-1mm} = \hspace{-1mm} \mathop{ \max } \limits_{1 \leq i < j \leq n} | g^{(2)}(i,j) |, \label{eq:relationmuandg2}
\end{eqnarray}
where $(a)$ is because the ensemble average of a random variable $\mathbf{y}$ approaches the mean value of its $m$ measurements for a large $m$. The contrast of speckle corresponding to the $i$-th column of $\mathbf{\Phi}$ is defined as 
$C( \phi_i )  =  {  ( \phi_i - \langle \phi_i \rangle )^T ( \phi_i - \langle \phi_i \rangle ) } /{   ( m \langle \phi_i \rangle \langle \phi_i \rangle ) }$. In particular, $C( \phi_i ) =1$ in GI because the probability of intensity of speckle obeys the negative exponential distribution. 

\begin{figure}[t]
\centering
\includegraphics[width = \linewidth]{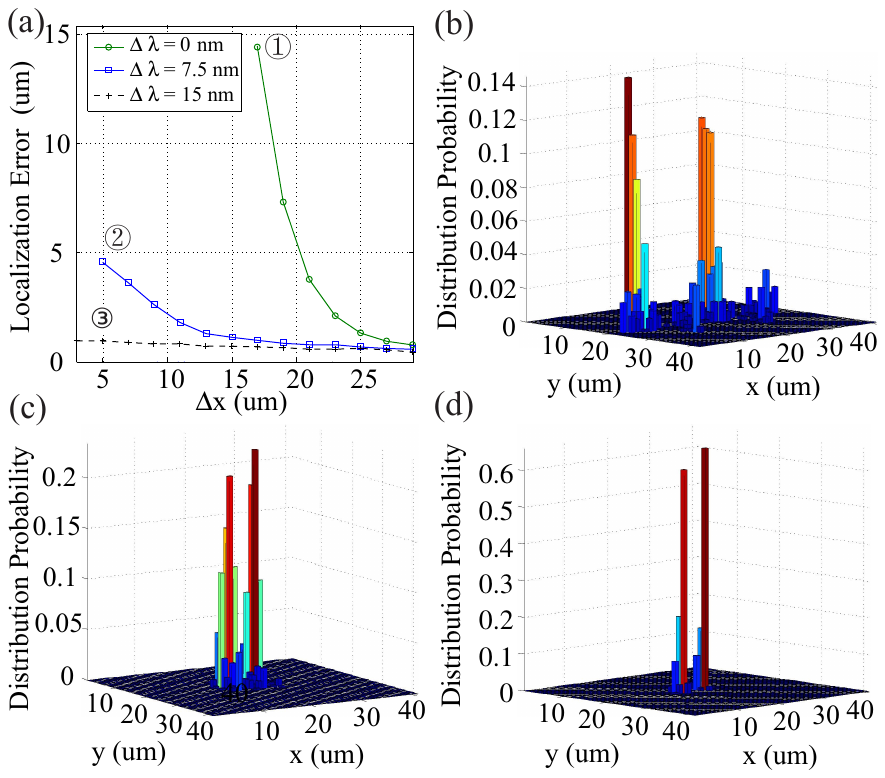}
\vspace{-2mm}
\caption{Localization error of GISC camera. (a) The average localization error as a function of distance ($\Delta \mathbf{x}$, $\Delta \lambda $) in high-dimensional light-field space. The spectral resolution of GISC camera is about $19$ nm. (b)--(d) The distribution probability of positions corresponding to the labels $1,2,3$ in (a).  }
\label{Simulation of GISC}
\vspace{-3mm}
\end{figure}

\begin{figure}[t]
\centering
\includegraphics[width=\linewidth]{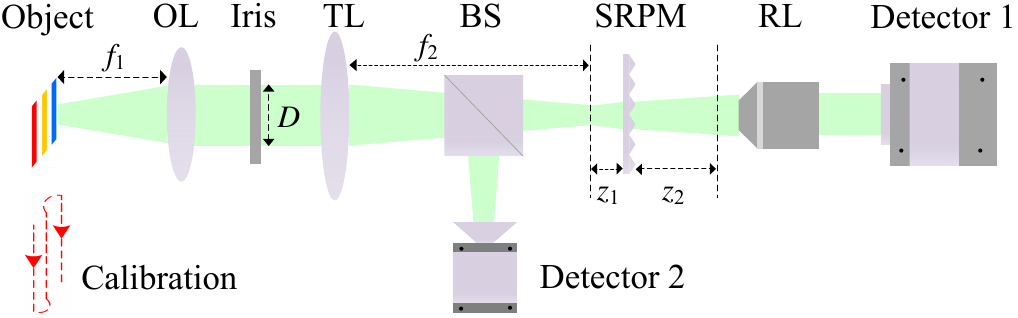}
\vspace{-3mm}
\caption{Experimental setup of GISC camera. A multicolour object through a traditional imaging system, which consists of an objective lens (OL) with $f_1 = 50$ mm and $D=1.48$ mm, and tube lens (TL) with $f_2 = 300$ mm, is modulated into a speckle image by spatial random phase modulator (SRPM), where $z_1=5$ mm and $z_2=40$ mm. 
Speckle image is expanded by relay lens (RL) to match the pixel size of detector $1$. For comparison, the diffraction limited image through the traditional imaging system is recorded on detector $2$ by another optical path via beam splitter (BS).  }
\label{setup}
\vspace{-3mm} 
\end{figure}

~\eqref{eq:relationmuandg2} shows that $\mu( \mathbf{\Psi} )$ is equal to the maximum value of $g^{(2)}(i,j)$, where $g^{(2)}(i,j)$ denotes the correlation of system responses of any two points in high-dimensional light-field space. 
In CS, $\mu < 1/(2K-1)$ is sufficient to guarantee the exact recovery of any $K$-sparse signal by the orthogonal matching pursuit (OMP) algorithm in the noiseless case~\cite{tropp2004greed}. Consequently, the nearest distance ($\Delta \mathbf{x}^{K}_{min}, \Delta \lambda^{K}_{min}$) among $K$ points in high-dimensional light-field space to guarantee the exact recovery of any $K$-sparse signal is given by
\begin{eqnarray}
\hspace{-.5mm}\exp \hspace{-1mm}\left [ { \hspace{-.75mm} - \hspace{-.75mm}  \left( \hspace{-.25mm}  \frac{ 2 \pi ( n \hspace{-0.25mm} -\hspace{-0.25mm}  1 \hspace{-0.25mm} )  \omega \Delta \lambda^{K}_{min} }{ {\overline{ \lambda }} ^2} \hspace{-.5mm} \right)^2 }\right ] \hspace{-1mm} \left( \hspace{-.5mm} \frac{ 2 \hspace{-0.25mm}  J_1 \hspace{-0.25mm} ( \hspace{-0.25mm}  \frac{ \pi D | \Delta \mathbf{x}^{K}_{min} | }{ \overline{ \lambda } f_1 }  \hspace{-0.25mm} ) } {  \frac{ \pi D | \Delta \mathbf{x}^{K}_{min}| } {  \overline{ \lambda } f_1 }} \hspace{-.5mm} \right) ^2 \hspace{-.75mm} <  \hspace{-.75mm} \frac{1}{2K \hspace{-0.25mm} -  \hspace{-0.25mm} 1}, \label{eq:condition}
\end{eqnarray}
see Supplementary material for more details on derivation of~\eqref{eq:condition}.
Clearly~\eqref{eq:condition} reveals that the statistical resolution~\cite{helstrom1964detection} of GISC camera heavily relies on the sparsity level $K$ of signals. Moreover, one can interpret from~\eqref{eq:condition} that the resolved spatial resolution $\Delta \mathbf{x}$ could be arbitrarily small in principle, as long as the spectral difference $\Delta \lambda_{min}$ of object is large enough to guarantee~\eqref{eq:condition}. However, due to i) the limitations of system parameters in~\eqref{eq:condition} and ii) limited number of samples and detection signal-to-noise ratio (SNR), the spatial resolution of GISC camera cannot be improved infinitely in practical applications.

In conventional ``point-to-point" imaging mode, the imaging resolution is determined by the sampling frequency and accuracy of ``point-to-point'' measurement that is restricted to the system parameters. Once the sampling frequency exceeds Nyquist limit, finer sampling inside the cell in which the thermal light field is partially coherent in frequency domain and spatial domain provides few further improvements on resolution. 
Different from the ``point-to-point" imaging mode, the statistical resolution of GISC camera is determined by degrees of freedom in objects' spatial-spectral distribution (i.e., $K$) and mutual coherence of sampling matrix if only the number of samples is large enough. Spatial sampling interval could be arbitrarily small in principle as long as the differences between the adjacent voxels in other degrees of freedom of light field, e.g., spectrum, is large enough to ensure the sampling matrix with small mutual coherence, which could guarantee the $K$-sparse signal exactly recovery. In addition, the normalized second-order correlation function $g_i^{(2)} \leq 1$ for each independent dimension, e.g., spatial or spectral dimension, of light field, and the global normalized second-order correlation function $g^{(2)}$ of the imaging system is the product of all normalized second-order correlation function $g_i^{(2)}$.
Thus, the larger degrees of freedom of light field in the imaging process and the smaller degrees of freedom in objects' spatial-spectral distribution are, the higher statistical resolution of GISC camera will be. The relationship between imaging resolution, degrees of freedom of light field, and degrees of freedom in objects' spatial-spectral distribution is illustrated in~\eqref{eq:condition}.

Consider the parameters of GISC camera: $f_1 = 50$ mm, $D = 1.48$ mm, $n = 1.53$, $\omega = 3.5$ um, $\zeta = 12$ um, and $\overline{\lambda} = 540$ nm. The resolution limit for $K = 2, 3, 5, 7$ in high-dimensional light-field space is shown in Figure.~\ref{fig:b}, where Rayleigh's spatial resolution criterion for the objective lens is $ |\Delta \mathbf{x}_s| = 1.22\overline{ \lambda} f_1 / D \approx 22.2$ um, and the conventional spectral resolution of GISC camera is $\Delta \lambda_s = 20.8$ nm. It can be observed that the statistical spatial resolution of GISC camera can be enhanced as the discernibility in spectral dimension increases.
For fixed $\Delta \lambda$, the statistical spatial resolution $\Delta \mathbf{x}$ gets worse as the sparsity $K$ goes large. When object is not sparse in real space, namely the sparsity $K$ of object in real space is very large, the statistical spatial resolution of GISC camera without spectral difference would approach the classical Rayleigh’s limit.

\section{Simulations and Experiments}

\begin{figure*}[t] 
\centering
\hspace{-5mm}\subfigure[~~Resolved spatial distance as a function of the spectral gap $\Delta \lambda$]
{\includegraphics[width = 80mm, height= 65mm] {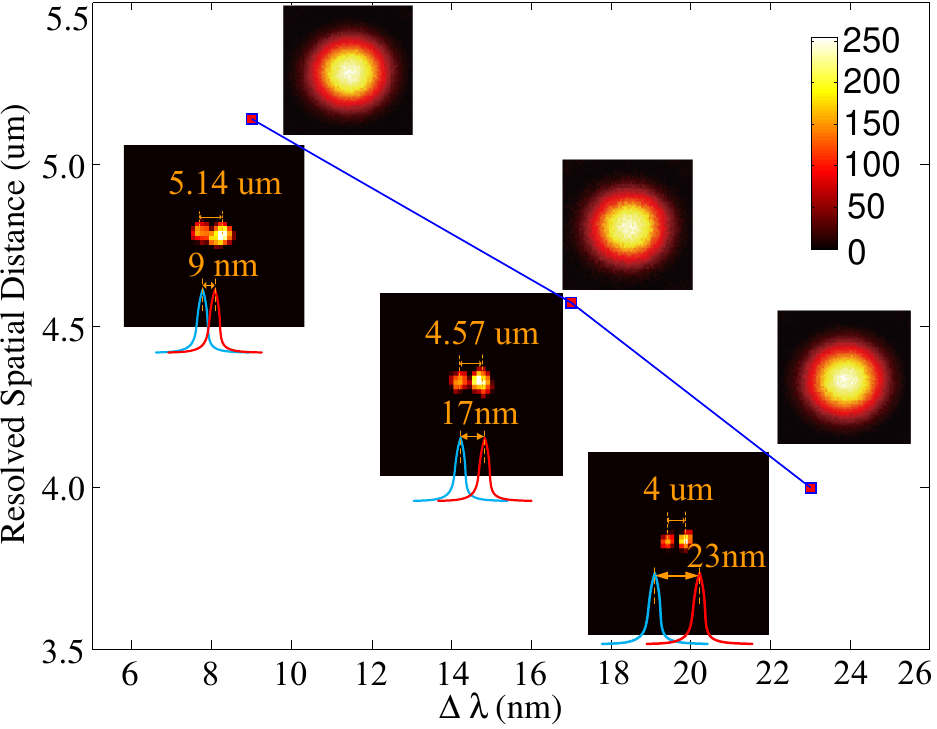}
\label{fig:lambda} }
\hspace{5mm}\subfigure[~~Resolved spatial distance as a function of the polarization state's difference $\Delta \theta$]
{\includegraphics[width = 80mm, height= 65mm] {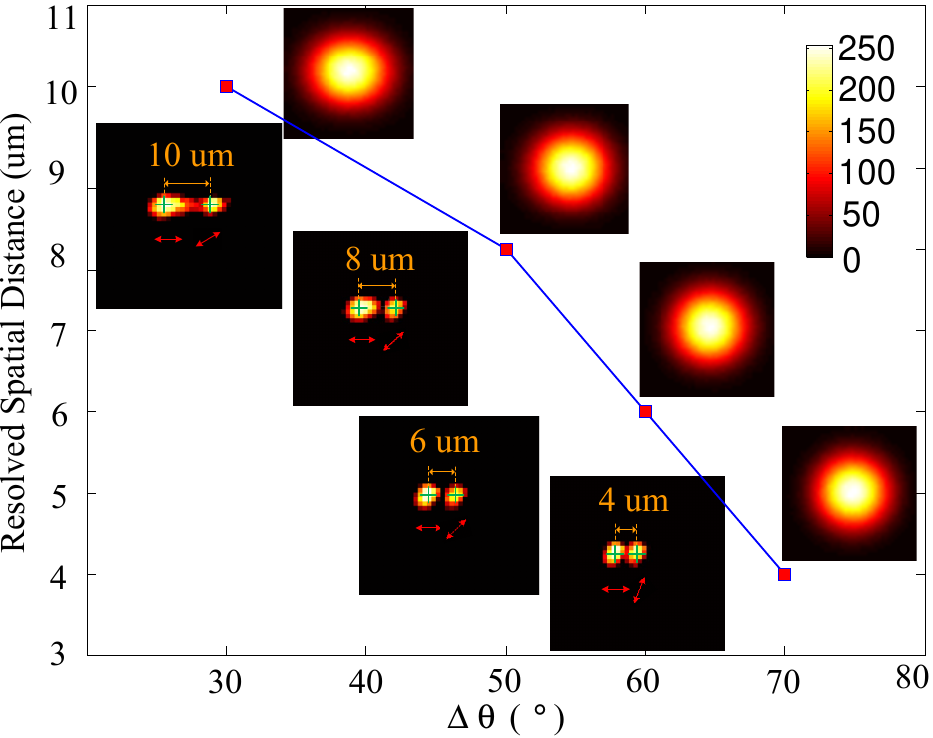}
\label{fig:polarization} }
\vspace{-3mm}
\caption{Experimental results for the relationship between resolved spatial distance and discernibility in high-dimensional light-field space. (a) Resolved spatial distance as a function of spectral gap $\Delta \lambda$. (b) Resolved spatial distance as a function of the polarization state's difference $\Delta \theta$. The upper ones in figure are diffraction limited images, and the bottom ones are reconstructed images with different spatial distance and different spectral gap or different polarization state.}
\vspace{-3mm}
\label{fig:ExperimentalResulttwopoints}
\end{figure*} 

\begin{figure}[t]
\centering
\includegraphics[width=0.95\linewidth]{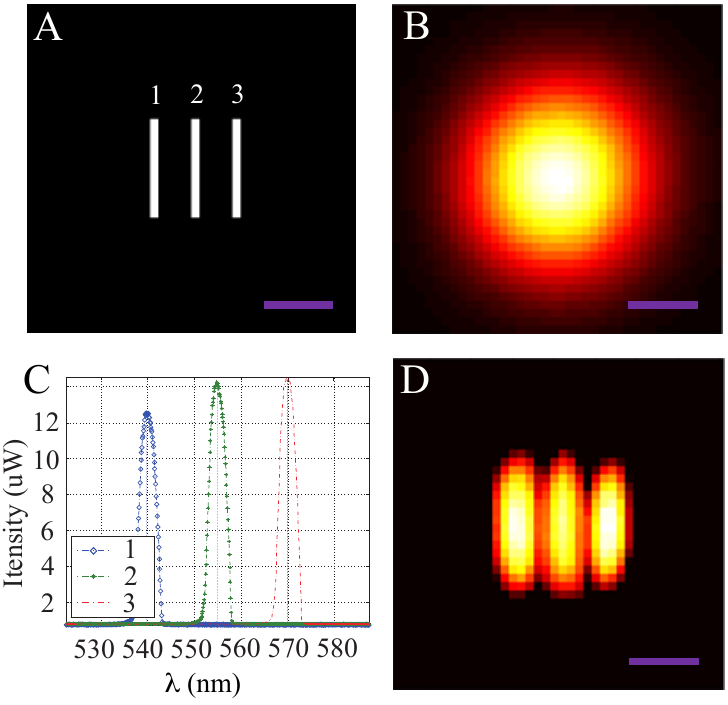}
\vspace{-3mm}
\caption{Experimental results of three slits. (A) Object consists of three slits, the spatial distance between each slit is $5$ um; the purple scale bar is $10$ um. (B) Diffraction limited image. (C) Spectral distribution of each slit, and the spectral gap between each slit is $15$ nm. (D) Reconstructed image.}
\label{fig:Experimentalresults}
\vspace{-3mm}
\end{figure}

\subsection{Simulation Results}
In order to verify the analytical result, we perform simulations. In the simulation, the parameters of GISC camera are set as $f_1 = 50$ mm, $f_2 = 300$ mm,  $D = 4$ mm, $n = 1.53$, $\omega = 3.5$ um, $\zeta = 12$ um, $z_1=5$ mm, and $z_2=40$ mm. Firstly, the sampling matrix whose columns consist of speckle patterns is created, where the speckle patterns are simulatively obtained corresponding to a  $64$ um $\times$ $64$ um field of view (FoV) and wavelength range from $530$ nm to $560$ nm with interval $15$ nm; see Supplementary material for more details about the simulation. Secondly, a resolution test target is constructed (Figure~\ref{fig:SimulationResults}.A), and the spatial distances between each slit are respectively $6$ um, $5$ um, $4$ um, and $3$ um. To imitate the diffraction limit of the objective lens, we convolute the image of resolution test target with a point-spread function (PSF)~\cite{born2013principles}, $\left( 2 \hspace{-0.25mm}  J_1 \hspace{-0.25mm} ( \hspace{-0.25mm}  \frac{ \pi D | \mathbf{x} | }{ \lambda  f_1 }  \hspace{-0.25mm} ) / {  \frac{ \pi D | \mathbf{x}| } { \lambda f_1 }} \hspace{-.5mm} \right) ^2$, of the objective lens to get the diffraction limited image (Figure~\ref{fig:SimulationResults}.B), where the full width half maximum (FWHM) of PSF is about $8$ um at wavelength $530$ nm. The sampling signals $\mathbf{Y}$ of GISC camera for the resolution test target with or without spectral difference are simulated via~\eqref{sample}. 
By solving problem~\ref{equation: measurement} with the gradient projection for sparse reconstruction (GPSR) algorithm~\cite{figueiredo2007gradient}, the reconstructed result of resolution test target without spectral difference shows that the spatial distance below $5$ um is hard to resolve (Figure~\ref{fig:SimulationResults}.D). Nevertheless, when resolution test target has spectral difference, where the labels $1,2,3$ on the resolution test target have different wavelength (Figure~\ref{fig:SimulationResults}.C),  the spatial distance as low as $3$ um in the reconstructed result can be recognized (Figure~\ref{fig:SimulationResults}.E). As observed in Figure~\ref{fig:SimulationResults}.F, the spatial resolution of GISC camera is obviously improved with the spectral difference. 

The localization error of GISC camera in high-dimensional light-field space is also investigated by simulations. In the simulation, we construct a serials of objects that consist of two point sources with different spatial distance ($\Delta \mathbf{x}$ = $3$ um, $5$ um, $\cdots$, $29$ um) and different spectral gap ($\Delta \lambda$ = $0$ nm, $7.5$ nm, $15$ nm), where the spectral resolution of GISC camera is about $19$ nm. The sampling signals $\mathbf{Y}$ are generated through~\eqref{sample} with an experimental sampling matrix $\mathbf{\Phi}$. For each distance ($\Delta \mathbf{x} , \Delta \lambda$), we perform $500$ independent trials by randomly changing the spatial positions of two point sources with the fixed distance. In each trial, the OMP algorithm is used to recover the points, and the Euclidean distance between the recovered points and truth points is calculated, which presents the localization error. By averaging the values over $500$ independent trials, we plot the results as a function of distance ($\Delta \mathbf{x}$, $\Delta \lambda $) in Figure~\ref{Simulation of GISC}(a), and the distribution probability of recovered points in Figure~\ref{Simulation of GISC}(b)-(d). It can be observed that the simulation results well match those in our theoretical analysis. In particular, the average localization error decreases as the distance ($\Delta \mathbf{x}$, $\Delta \lambda $) in high-dimensional light-field space increases, which clearly demonstrates that the discernibility in high-dimensional light-field space indeed improves the statistical spatial resolution of GISC camera.

\subsection{Experimental results}
The experimental setup of GISC camera is illustrated schematically in Figure~\ref{setup}. It has two processes to recover the high-dimensional image of object: 1) pre-determined calibration process and 2) imaging process. In the pre-determined calibration process, a monochrome point source is fabricated by using an optical fiber coupling the monochrome light from a wide-band white source (Newport, Arc lamp $66902$). The incident light wavelength varies from $540$ nm to $ 615$ nm with interval $\Delta\lambda = 7.5$ nm. The point source shifts at a step size $\Delta\mathbf{x} = 1$ um in the $40$ um $\times$ $ 40$ um FoV on the object plane, and corresponding speckle patterns are recorded to form each column of sampling matrix $\mathbf{\Phi}$, respectively. In the imaging process, a multicolor object is projected on the image plane through the traditional imaging system, and is further modulated into a speckle image by SRPM (Thorlab, DGUV 10--1500). A relay lens (Olympus, UPlanSApo $4\times$) is used to match the pixel size of detector $1$ (Andor, iKon--M, pixel size of $13$ um) and the average size of the pre-determined speckle patterns. For comparison, the diffraction limited image due to the objective lens is recorded on detector $2$ (Allied Vision, Stingray F--$504$B) at another optical path split by a $10/90$ beam splitter. Rayleigh's spatial resolution criterion on the object plane is about $22.2$ um at wavelength $540$ nm.

In the experiments, two point sources with different spatial distance $\Delta \mathbf{x}$, different spectral gap $\Delta \lambda$ and different polarization state $\Delta \theta$ are constructed. In the first experiment, two point sources with three different spatial distances $4$ um, $4.57$ um and $5.14$ um and corresponding spectral gaps $23$ nm, $17$ nm and $9$ nm are constructed. The GPSR algorithm is adopted to solve problem~\ref{equation: measurement}. It can be observed in Figure~\ref{fig:lambda} that the resolved spatial distance of GISC camera is improved by increasing spectral difference. 
In the second experiment, two point sources with different linear polarization state are fabricated. Specifically, the spatial distances between two point sources are $4$ um, $6$ um, $8$ um and $10$ um, respectively, and the difference of polarized angle between two point sources varies from $0^\circ$ to $180^\circ$. By rotating a polarizer to four different detections of polarization $\theta$ ($0^\circ$, $45^\circ$, $90^\circ$, and $135^\circ$) in front of detector $1$, a polarized detector is built to record the sampling signals ($\mathbf{Y}_{0^\circ}$, $\mathbf{Y}_{45^\circ}$, $\mathbf{Y}_{90^\circ}$, $\mathbf{Y}_{135^\circ}$).  By solving problem~\ref{equation: measurement} with GPSR algorithm, the object with four polarized detections ($\mathbf{X}_{0^\circ}$, $\mathbf{X}_{45^\circ}$, $\mathbf{X}_{90^\circ}$, $\mathbf{X}_{135^\circ}$) are reconstructed, and the intensity of object is achieved by summing $\mathbf{X}_{0^\circ}$ and $\mathbf{X}_{90^\circ}$. Figure~\ref{fig:polarization} presents the resolved spatial distance as a function of the polarization state's difference $ \Delta \theta $. One can observe that the resolved spatial distance is also improved by increasing the difference of the polarization state. Overall, the resolved spatial distance of GISC camera can be promoted by enhancing the discernibility in high-dimensional light-field space. It is worth mentioning that the best resolved spatial distance of GISC camera can achieve $4$ um at spectral gap $23$ nm or polarization state's difference $70^{\circ}$ in the experiment, which is over five folds of Rayleigh's spatial resolution criterion. 

Figure~\ref{fig:Experimentalresults} presents the experimental result of three slits. The spatial distance and spectral gap between each slit are $5$ um and $15$ nm, respectively. The spectral width of each slit is about $4.5$ nm. Figure~\ref{fig:Experimentalresults} (B) and (D) are the corresponding diffraction limited image and the reconstructed super-resolution image. It can be observed that three slits are resolved even when the spatial distance $5$ um between each slit is far smaller than the diffraction limit $22.2$ um, which implies that over four folds of Rayleigh's spatial resolution criterion is achieved in our experiment due to the discernibility in the ($\mathbf{x}, \lambda$) light-field space.

\section{Discussion and conclusion}
In this paper, we demonstrated both theoretically and experimentally that, 
contrary to eliminating the chromatic aberration relative to different wavelength in traditional camera,
the discrepant information of light field irradiated from object in the non-spatial dimension can be exploited to break Rayleigh's spatial resolution criterion of GISC camera. While ``Rayleigh's curse'' induced by the diffraction effect still exists when resolving two closely located points in high-dimensional light-field space, it could be avoided in the spatial dimension as long as the two points are distinguishable in other dimensions, such as the spectrum and polarization. For conventional imaging, the sampling constraint is Nyquist sampling theorem, namely, finer sampling exceeding Nyquist limit provides few further improvements on resolution. However, for GISC camera, there is no constraint on sampling as long as~\eqref{eq:condition} is satisfied.

Compared with current physical super-resolution techniques, GISC camera fully utilizes the imaging channel capability of imaging system, and has high information acquisition efficiency. Therefore, it can realize the wide-field super-resolution imaging in one snapshot with the detection sensitivity and SNR guaranteed. However, as a computational imaging technique, the super-resolution ability of GISC camera heavily relies on recovery algorithms, and the computational cost for a large-scale sampling matrix is prohibitively high. There is a large room for developing high-efficient recovery algorithms by integrating more prior information~\cite{Scarlett2013CompressedIToSP}, novel processing methods for sampling matrix~\cite{tong2020preconditioned}, and new recovery approaches to reduce the computational cost~\cite{Palangi2016DistributedIToSP}. Moreover, by optimizing the SRPM for higher imaging efficiency~\cite{hu2019optimization} and higher detection SNR~\cite{liu2020spectral}, the resolution of GISC camera could be further improved. 

Recently, GISC nanoscopy with $80$ nm spatial resolution in single frame has been demonstrated by utilizing the sparsity of fluorescence emitters~\cite{li2019single}. As validated in this paper, GISC nanoscopy exploiting the discernibility in high-dimensional light-field space will further promote its performance in breaking the classical Rayleigh's limit. Together with the rapid development in fields of signal processing for sparse recovery~\cite{Marques2019ReviewIA,Friedland2014tcs}, light-field optimization based on metasurfaces~\cite{Aieta2015Multiwavelength}, megapixels and polarized detection technologies~\cite{Canon2015Canon,yamazaki2016four}, GISC camera could be widely applied in the multicolor fluorescence microscopy~\cite{bates2007multicolor}, stimulated Raman scattering microscopy~\cite{freudiger2008label}, polarization microscopy~\cite{Zhanghao2017SuperJoIOHS}, and the wavelength and polarized astronomy~\cite{wilking1980wavelength}. 

\vspace{3mm}




\ifthenelse{\equal{\journalref}{aop}}{%
\section*{Author Biographies}
\begingroup
\setlength\intextsep{0pt}
\begin{minipage}[t][6.3cm][t]{1.0\textwidth} 
  \begin{wrapfigure}{L}{0.25\textwidth}
    \includegraphics[width=0.25\textwidth]{john_smith.eps}
  \end{wrapfigure}
  \noindent
  {\bfseries John Smith} received his BSc (Mathematics) in 2000 from The University of Maryland. His research interests include lasers and optics.
\end{minipage}
\begin{minipage}{1.0\textwidth}
  \begin{wrapfigure}{L}{0.25\textwidth}
    \includegraphics[width=0.25\textwidth]{alice_smith.eps}
  \end{wrapfigure}
  \noindent
  {\bfseries Alice Smith} also received her BSc (Mathematics) in 2000 from The University of Maryland. Her research interests also include lasers and optics.
\end{minipage}
\endgroup
}{}

\end{document}